\documentstyle[12pt]{article}
\begin{document}
\noindent{\parbox[t]{2in}{\footnotesize To appear in the proceedings of NATO
ASI /Euroconference
{\bf Formation and Interactions of Topological Defects},
A. C. Davis and R. N. Brandenberger, eds. (Plenum, in press)}
\hfill
\parbox[t]{1.5in}{Los Alamos preprint LAUR 95-170\\
Newton Institute preprint NI 94044}}
\baselineskip=18pt plus 0.2pt minus 0.2pt
\lineskip=18pt plus 0.2pt minus 0.2pt
\vspace*{0.8in}

\noindent {\bf COSMOLOGICAL EXPERIMENTS IN SUPERFLUIDS}\\
\noindent {\bf AND SUPERCONDUCTORS}
\vspace*{0.55in}

\noindent \hspace*{1.00in} W. H. Zurek\\
\hspace*{1.00in} Theoretical Astrophysics, T-6, MS B288\\
\hspace*{1.00in} Los Alamos National Laboratory, Los Alamos, NM  87545
\vspace*{0.4in}

\baselineskip=14pt
\noindent {{\bf Abstract:}  Evolution of the order parameter in
condensed matter analogues of cosmological phase transitions is
discussed.  It is shown that the density of the frozen-out topological
defects is set by the competition between the quench rate -- the rate
at which the phase transition is taking place -- and the relaxation
rate of the order parameter.  More specifically, the characteristic
domain size which determines the typical distance separating
topological defects in the new broken symmetry phase (and, therefore,
their density) is determined by the correlation length at the instant
at which the relaxation timescale of the order parameter is equal to
the time from the phase transition.  This scenario shares with the
Kibble mechanism the idea that topological defects will appear ``in
between'' domains with independently chosen broken symmetry
vacuum. However, it differs from the original proposal in estimating
the size of such domains through the non-equilibrium aspects of the
transition (quench rate), rather than through the Ginzburg temperature
at which thermally activated symmetry restoration can still occur in
the correlation - length sized volumes of the broken symmetry phase.
This scenario can be employed to analyze recent superfluid quench experiments
carried out in bulk He$^4$ to study the analogue of the
``cosmological'' prediction of significant vortex line production.  It
can be also applied to superfluid quenches in annular geometry, as
well as to the rapid phase transition from the normal metal to
superconductor, where the symmetry breaking occurs in the order
parameter with the local (rather than a global) gauge.  Cosmological
implications of the revised defect formation scenario with the
critical domain size set by the freeze-out time rather than by the
Ginzburg temperature are also briefly considered.}

\baselineskip=18pt plus 0.2pt minus 0.2pt
\bigskip
\bigskip

\noindent {\bf 1.\quad INTRODUCTION}\\

Expansion of the Universe following the Planck-era ``Big Bang''
inevitably leads to the decrease of temperature of the primordial
fireball.  This is thought to precipitate phase transitions which
transform the vacuum from the ``false'' symmetric, high temperature
phase to the low temperature broken symmetry ``true'' vacuum with the
structure which defines ``low energy physics'' accessible to us in
(high energy!) experiments.  As the Universe undergoes phase
transitions, the selection of the low temperature broken symmetry
phase can only occur locally, within the causally correlated regions.
It was noted by Zeldovich$^1$ and his co-workers and especially by
Kibble,$^2$ that this symmetry breaking process may leave relics of
the high energy phase -- islands of the symmetric ``false'' vacuum --
which will be trapped by the topologically stable configurations of
the broken symmetry phase.  Such topological defects would be massive
and would therefore have observable consequences for the structures
forming within the Universe as well as for the cosmic microwave
background or for the evolution of the Universe as a whole.

Three principal kinds of topological defects$^3$ are distinguished by
their dimensionality.  Monopoles are pointlike, and a disaster from
the cosmological point of view.  Membranes are two-dimensional, and
(almost certainly) also a disaster: They would cause unacceptably
large distortions of the cosmic microwave background.  One -
dimensional cosmic strings are by contrast a source of density
perturbations which are still under investigation as a possible seed
of structure formation.

Symmetry breaking phase transitions which occur in condensed matter
physics are described by theories which are formally identical to
those involved in the cosmological context, but have one crucial
advantage: they can be studied in the laboratory.  With this in mind,
almost exactly a decade ago I have suggested$^{4,5}$ that the
cosmological mechanism for defect formation can be studied
experimentally in the condensed matter context.  The aim of this paper
is to review this idea and to assess the experimental progress in
implementing this cosmological scenario in the various condensed
matter systems as well as to sketch possible directions for the
future research.

This is an excellent time to undertake such a reassessment; the first
realization of ``cosmological experiments'' has been
accomplished a few years ago by Bernard Yurke
and his colleagues in liquid crystals$^{6,7}$.  Even more recent
exciting development is the experiment carried out by Peter McClintock
and his colleagues$^{8,9}$ who have implemented the original proposal
by studying the superfluid transition in He$^4$.

Liquid crystal experiments demonstrated that copious production of
topological defects does indeed happen.$^{6,7,10}$ But as the phase
transition is of the first order, the interesting dynamics which takes
place in the second order (Landau-Ginzburg like) phase transitions
which are relevant to cosmology cannot be directly
studied.  The size of the characteristic domains (and the density of
defects) is then set instead by the nucleation process.  By contrast,
liquid He$^4$ becomes superfluid without nucleation.  Therefore, while
both liquid crystals and superfluids are of great interest, one might
argue that the $\lambda$-transition into the superfluid allows one to
address questions which cannot be posed in the liquid crystals
context.  (On the other hand, topological defects can be seen directly
in liquid crystals, which means that the reverse of the above
assertion is also true!)

Both liquid crystals and superfluids are described by an order
parameter with a global gauge symmetry.  While global gauge field
theories may be relevant to cosmological models, they seem to be an
exception rather than the rule: Theories with local gauge symmetry are
therefore even more interesting as an analogue of the cosmological
phase transitions.  Superconductors offer an obvious condensed matter
example, and we shall also discuss the possibility of implementing
``cosmological'' quenches in this alternative low temperature setting.

Defect formation scenario in course of the rapid phase transitions is
based on two assumptions.  The first assumption is of the qualitative
nature: It asserts that regions of the broken symmetry phase which are
causally disconnected must select the new low temperature phase
independently. As a result, when a symmetry-breaking phase transition
with a non-trivial homotopy group occurs simultaneously in a
sufficiently large volume, topological defects will appear with
\underline{some} density.  The second assumption is of the
quantitative nature: It involves specifying the process
responsible for the causal propagation of ``signals'' which allows the
choice of the new vacuum to occur in a coordinated (rather than
independent) fashion.  It leads to prediction of the density of
topological defects.

Both assumptions are of course necessary$^2$, but while the first one
is straightforward and (with the benefit of hindsight) hard to argue
with, the second one requires much more specific physical input.  In
the first order phase transitions the process which is responsible for
the appearance of the new phase is nucleation: Small regions of the
medium undergo thermal activation which takes them over the potential
barrier separating ``false'' and ``true'' vacua.  As a result, bubbles
of a certain (critical) size appear and form seeds of the new phase.
Eventually, through growth and coalescence of these bubbles new phase
replaces the old one.

The original discussion of the scenario for defect formation appealed
to a similar idea$^2$: It was thought that thermally activated
transitions between the correlation-sized volumes of the new broken
symmetry phase which are still possible well below the critical
temperature T$_C$ determine the initial density of the topological
defects.  Such transitions may occur down to the so-called Ginzburg
temperature T$_G$ (T$_G <$ T$_C$), at which the (free) energy barrier
becomes prohibitively large for correlation-length sized thermal
fluctuations.  If this were indeed the case, density of defects would
be set by the correlation length at the Ginzburg temperature.

One of the key predictions of the original papers$^{4,5}$ on
``cosmological'' phase transitions in superfluids was that this
thermally activated process \underline{does not} decide the density of
defects: It was conjectured that the corresponding transitions are too
local to result in ``global'' structures such as strings or domain
walls.  Instead it was proposed that the characteristic correlations
length is set by the dynamics of the order parameter in the vicinity
of the critical temperature T$_C$.

Both in the superfluid (or, more generally, in the condensed matter)
context and in the cosmological phase transitions selection of the new
vacuum cannot be dynamically coordinated over regions larger than the
size of the ``relevant causal horizon.''  In superfluids (as well as
in the case of other second-order phase transitions with a
non-conserved order parameter) the velocity with which perturbations
of the order parameter can propagate yields a natural ``sonic
horizon.''  I shall show that such considerations appear to lead to
reasonable estimates of the density of topological defects in He$^4$
superfluid quench experiments$^{8,9}$, and demonstrate that this
estimate differs from the one based on thermal activation and Ginzburg
temperature.

The above ``nonequilibrium'' scenario$^{4,5}$ represents a significant
change of the point of view, and yields a prediction for the density
of defects which is rather different from the ``equilibrium'' estimate
based on the Ginzburg temperature.$^2$  The aim of the rest of this paper
is to physically motivate, describe, and investigate consequences of
this ``freeze-out'' scenario on the example of superfluids.  It is
hoped that this discussion can be then generalized to the example of
other phase transitions described by the non-conserved order parameter
with global gauge symmetry. I shall also discuss the more complicated
case of superconducting phase transition (where the broken symmetry
phase is described by a locally gauge invariant theory) and comment on
the implications of these considerations for the cosmological phase
transition scenarios.

\bigskip
\bigskip
\noindent
{\bf 2. SYMMETRY BREAKING IN SUPERFLUID HELIUM, SUPERCONDUCTORS, AND
IN THE EARLY UNIVERSE}\\

The aim of this section is to review some of the equilibrium
properties of the systems which will serve as condensed matter
analogues of the cosmological phase transitions.  We shall carry out
our discussion starting with the superfluid He$^4$, go on to discuss
superconductors, and finish with a brief overview of symmetry breaking
in the field theories relevant for cosmological phase transitions.
Many of the prospective readers of this paper may be used to the order
-- common in the cosmology/particle astrophysics presentations --
which covers the same ground, but in the opposite directions, and with
a complementary emphasis.  I have adopted this order starting with the
low temperature phase transitions for several reasons: to begin with,
it is meant to emphasize that the most accessible testing grounds for
the ideas we shall be considering are in the simplest low temperature
systems.  (This belief has been strengthened by the recent experiments
of McClintock and his co-workers.$^{8,9}$) Moreover, this sequence
reflects the original flow of ideas (where the condensed matter
analogues were used to elucidate spontaneous symmetry breaking in the
field-theoretic context).  Finally, this order correlates with the
degree of confidence we can have also in the non-equilibrium aspects
of various examples of the phase transitions, and especially in the
resulting scenarios of topological defect formation.

\bigskip

\noindent {\bf Superfluid Helium}\\
\indent Superfluid forms in the low temperature (T$ < 2.18 ^{\circ}K$),
moderate pressure (p$<
25$ atm) corner of the He$^4$ phase diagram.  Various manifestations
of superfluidity (like flow with negligible friction, persistent
currents, etc.) are well-documented in the literature.$^{11}$ They can
be accommodated and explained the context of several phenomenological
models, which tend to emphasize various properties of superfluid
helium.  Thus, two-fluid model introduced by Tisza regards superfluid
He$^4$ as a mixture of two components -- superfluid and normal.  The
fraction of the fluid which is either ``normal'' or ``super'' depends
on the distance from the (pressure dependent) temperature
T$_{\lambda}$ at which transition to the normal He$^4$ liquid occurs,
which is known as the ``$\lambda$ line'' (because of the asymmetric
form of the specific heat near T$_{\lambda}$).

While normal He$^4$ is simply another liquified noble gas, properties
of the superfluid fraction can be qualitatively understood when it is
regarded as a Bose condensate of He$^4$ atoms.  In particular, in
addition to the normal ``first sound'' (in which density perturbations
propagate as in the air) superfluid He$^4$ can carry (albeit much more
slowly) the \underline{second sound} (where the relative densities of
the super and normal components are perturbed, but in such a way that
the total density remains constant).

Phonons of the second sound are not the only new excitations in the
superfluid phase: Rotons -- massive excitations -- also appear below
T$_{\lambda}$.  Appearance of a new massless and a new massive
``particle'' in the broken symmetry phase below T$_{\lambda}$ is
strongly reminiscent of the Goldstone boson and of the Higgs particle
which should accompany breaking of the global symmetry during phase
transitions.

To pursue this (imperfect, as it turns out) analogy further we shall
rely on the Landau-Ginzburg theory of the second order phase
transitions.  There, the specific free energy of the system is given
in the vicinity of the phase transition by the analytic expression of
the form:
$$
F(\Psi) = \frac{\hbar^2}{2m} | {\vec \nabla} \Psi |^2 + \alpha
|\Psi|^2 + \frac{\beta}{2} |\Psi|^4 \ .  \eqno(1)
$$
Here $\Psi$ is the space-dependent \underline{order parameter}, an
abstract measure of the degree to which the symmetry in question has
been broken, while $\alpha$ and $\beta$ are parameters:
$$
\alpha = \alpha^{\prime} ~ (T-T_C) /T_C, \;\;\;\;\; \alpha^{\prime} > 0   \ ;
\eqno(2a)
$$
$$
\beta = {\rm const} > 0  \ . \eqno(2b)
$$
In addition to the two ``potential'' terms (which depend on the even
powers of the order parameter), Eq.~(1) contains the square of the
gradient, the ``kinetic energy'' contribution to the free energy.  The
mass $m$ is usually taken to be that of the He$^4$ atom, but is in
fact a parameter which is fixed by the normalization so that
$\left|\Psi (\vec{r})\right|^2$ yields the correct density of the
superfluid.

As the temperature drops below the critical $T_C = T_{\lambda}$, the
shape of the potential contribution to the free energy changes.
Instead of a single minimum in a disordered phase with $\Psi = 0$ one
now expects the field to have a typical amplitude given by:
$$
\sigma = \sqrt{-\alpha/\beta}  \ , \eqno(3)
$$
and a random phase.

In the application of Landau-Ginzburg theory to superfluid He$^4$ one
tends to regard the order parameter $\Psi$ as a \underline{wave
function of the Bose condensate}.  $\Psi$ is then a complex field, and
its instantaneous configurations need to be characterized by both its
amplitude and its phase as a function of position.  The simplest such
solution is of the form:
$$
\Psi = \sigma \exp (i\theta) \ ,  \eqno(4)
$$
where $\theta$ is constant.

To investigate more interesting (and more complicated) solutions it is
useful to rescale the free energy of Eq.~(1) in terms of $\sigma$ and
of the correlation length:
$$
\xi = \frac { \hbar } { \sqrt{2m |\alpha|}} \ .  \eqno(5)
$$
In superfluid He$^4$, well below T$_C$, the correlation length is of
the order of a few \AA ngstroms.  In terms of the new variables
$\varrho = r/\xi, \ \eta = \Psi/\sigma$ the condition for the stable
configuration of $\Psi$ (i.e., for a minimum of $F(\Psi)$) can be
expressed as:
$$
\bigtriangledown^2 \eta = (|\eta|^2 - 1 ) \eta  \ . \eqno(6)
$$
In addition to the trivial solution given by Eq.~(4) (i.e. $|\eta| =
1$) Eq.~(6) also admits axisymmetric solutions of the form:
$$
\eta = \psi (\varrho) \exp in \phi \ ,   \eqno(7)
$$
where ($\varrho, \phi, z$) are the cylindrical coordinates.  Here, $n$
must be a natural number (otherwise, $\eta$ would not be
single-valued).  The radial part of the physically relevant solution
is regular near the origin ($|\psi| \sim \varrho^n, \ \varrho \ll 1$)
and approaches equilibrium density at large distances ($|\psi|^2
\simeq 1- n^2/ \varrho ^2, \ \varrho \gg 1$).  The phase of the
complete solution is $\theta = n \phi$ on any $\varrho = {\rm const}
>0$ circle, but remains undefined along the singular $\varrho = 0$.

Since $\Psi$ is the wave function, the gradient of the phase gives the
local superfluid velocity;
$$
\vec{v_S} = \frac{\hbar}{m} \vec{\bigtriangledown} \theta (\vec{r})\;,
\eqno(8)
$$
where $m$ is the mass of the He$^4$ atom.  Therefore, the axially
symmetric solution of Eq.~(6) is a \underline{vortex line} with a core
of width given by the correlation length $\xi$, Eq.~(5).  The
superfluid circulates with a radius-dependent velocity given by:
$$
 | \vec{v_S} | = v_{\varphi} = \frac { \hbar } { m } \frac n r \
. \eqno(9)
$$
Inside the core a symmetric vacuum -- the normal fluid -- makes up for
the density deficit caused by the decrease of $|\Psi|^2$.  $n$ is
known as a winding number. For energetic reasons vortex lines with $n
>1$ tend to dissolve into vortices with $n =1$.

Existence of the vortex lines in the superfluid Helium 4 has been
postulated by Onsager and Feynman as the only means of introducing
rotation into the superfluid \underline{without} violating the
condition of the single-valuedness of the Bose condensate wavefunction
$\Psi$.  Their existence has been since confirmed and their properties
were carefully studied$^{11}$.  Seen from the vantage point of the
Landau-Ginzburg theory of superfluidity vortex lines are a perfect
example of a global topological defect$^3$.

The analog of the vortex line in field theories relevant in the
cosmological context is a cosmic string.  As it was noted by
Kibble,$^2$ strings form when the first homotopy group $\Pi_1 (G/H)$
-- where $G$ and $H$ are the symmetry groups before and after the
phase transition -- is nontrivial.  For superfluid helium this is
indeed the case, as $G/H = U(1)$, and $\Pi_1 (G/H) = Z$.  Superfluid
vortex line is an analogue of a \underline{global} string -- that is,
a string associated with the breaking of a global gauge symmetry.  The
alternative \underline{local} strings are like the flux lines in
superconductors.  We shall consider them below.

\bigskip
\noindent {\bf Superconductors}\\
\indent Landau-Ginzburg model of the second order phase transition is only a
qualitative
approximation for the superfluids, but it turns out to be a
quantitatively accurate mean field theory for superconductors$^{12}$.
It is based on the observation that the wave function of the Bose
condensate of Cooper pairs -- which is the relevant order parameter --
has a free energy density given by:
$$
F = \frac{1}{4m} |(-i\hbar \vec\bigtriangledown - \frac {2 e} c \vec A) \Psi
|^2
+ \alpha |\Psi|^2 + \frac{1}{2} \beta |\Psi|^4 + \frac {B^2}{8\pi }+
E_0 \ , \eqno(10)
$$
where $2m$ and $2e$ are a mass and a charge of a Cooper pair, and
where we have incorporated terms due to external magnetic field in the
constant $E_0$.  Equation (10) differs from the Eq.~(1) through the
presence of electromagnetic (gauge) fields -- vector potential
$\vec{A}$ enters into the kinetic term through the usual replacement;
$$
\vec{\bigtriangledown} \longrightarrow \vec{\bigtriangledown} \pm
\frac{2ie}{\hbar c} \vec{A} \ , \eqno (11)
$$
and the induction $\vec B$ is given by:
$$
\vec B = \vec{\bigtriangledown} \times \vec{A} \ .  \eqno(12)
$$

Symmetry breaking occurs below the phase transition temperature T$_C$
when the coefficient $\alpha(T) = \alpha^{\prime} (T-T_C)/T_C$ becomes
negative.  As in the superfluid He$^4$ the order parameter acquires a
finite vacuum expectation value, Eq.~(3), and an associated phase
$\theta$.

Quantized vortices in superconductors emerge in a manner analogous to
the vortex lines in superfluid.  To see this, let us consider a closed
loop in the real space.  Suppose that the phase of the broken symmetry
vacuum changes by 2$\pi$ as one follows the path along the loop.  In
the superconductor, the current is related to the gradient of the
phase through:
$$
\vec{J} = 2e |\Psi|^2 \vec v_S  \eqno(13)
$$
where the velocity of the Cooper pairs is given by:
$$
\hbar \vec{\bigtriangledown} \theta = 2m\vec{\upsilon_S} + 2e \vec{A}/c \;.
\eqno(14)
$$

With a few additional assumptions about the axisymmetry of the
solution one can then employ Eq.~(13) to calculate magnetic induction:
$$
B = \frac{\Phi_0}{2\pi \lambda^2 } K_0 (r/\lambda) \ . \eqno(15)
$$
Here $K_0$ is the zero - order Hankel function of imaginary argument
and $\Phi_0$ is the flux quantum;
$$
\Phi_0 = hc/2e\;,  \eqno(16)
$$
where $c$ is the speed of light, and equals to 2.07 $\cdot 10^{-7}$
gauss cm$^2$.

London penetration depth $\lambda$ is given by
$$
\lambda^2 = m c^2 / (8\pi e^2 n_C) \ , \eqno(17)
$$
with the equilibrium density of Cooper pairs:
$$
n_C = \sigma^2 = |\alpha|/\beta.  \eqno(18)
$$
Thus, the flow pattern, and with it $B$, die of exponentially on a
scale set by $\lambda$.  This scale on which electromagnetic
interactions fall off exponentially can be
either small or large compared to the correlation length of the order
parameter in superconductors:
$$
\xi^2 = \hbar/(4m \alpha) \ , \eqno(19)
$$
which is analogous to the superfluid correlation length, Eq.~(5).  The
value of $\xi$ determines the size of the core of a vortex -- size of
the region where there is no Bose condensate of Cooper pairs.

In a typical superconductor far below T$_C$ correlation length has
values of the order of $\xi_0 \cong 10^3 \AA$, two orders of magnitude
larger than the corresponding quantity in the superfluid.  Thus while
in the superfluid correlation length is of the order of the
interatomic spacing, in the superconductors it is at least two orders
of magnitude larger.  This is the main reason why the mean field
Landau-Ginzburg theory is quite accurate for superconductor, but only
qualitatively correct for superfluids.$^{11}$

The ratio of the penetration depth and of the correlation length does
not depend on temperature and is the fundamental parameter of the
theory, crucial for the existence of vortices.  This is because the
flow around the vortex has an inner radius given by $\xi$ and an outer
radius given by $\lambda$.  Thus, $\xi$ must be smaller than $\lambda$
if the vortex lines are to exist.  The exact condition turns out to
be:
$$
\kappa = \lambda/\xi > 1/\sqrt{2}\;.  \eqno(20)
$$
This is the condition which distinguishes between type I
superconductors, in which vortices are not found, and type II
superconductors, satisfying inequality (20), in which they can exist.

Superfluid vortex was an analogue of a global string.  Superconducting
vortices are analogous to local strings -- the solution I have briefly
sketched out above is similar to the Nielsen-Olesen string solution in
the field theories with local gauge.

\bigskip

\noindent {\bf Field Theory}\\
\indent Expressions (1) and (10) for the free energy of a superfluid or of
a superconductor have more general field-theoretic analogs.  Thus, one
can consider a complex field $\varphi$ with a Lagrangian (we adopt in
this section convenient units $\hbar = c = 1$) given by:
$$
L(\varphi) = (\partial_\mu \varphi^*)(\partial^{\mu} \varphi) - \alpha
\varphi^* \varphi -\frac{\beta}{2} (\varphi^* \varphi)^2\;.  \eqno(21)
$$
For $\alpha < 0, \ \beta>0$, the potential in (21) has a minimum when
the absolute value of $\varphi$ is given by $\sigma
=\sqrt{-\alpha/\beta}$ of Eq.~(3).  In addition to the topological
defects (which can be shown to exist through a discussion analogous to
our above derivation of vortex lines in superfluids), Eq.~(21) can be
used to demonstrate existence of massive and massless modes in the
broken symmetry vacuum.  To show this, one considers small
perturbations around the broken symmetry ground state (which can be
set for the purpose of this calculation to be completely real):
$$
\varphi(x) = \sigma + (u(x) + i \upsilon (x)) / \sqrt{2}\;.  \eqno(22)
$$
Substituting this into Eq.~(21) and ignoring constant terms one
recovers:
$$
L= \frac{1}{2} (\partial_\mu u)^2 + \frac{1}{2} (\partial_\mu
\upsilon)^2 - \beta \sigma ^2 u^2 - \frac{\beta}{\sqrt{2}} \sigma u
(u^2 + \upsilon^2) -\frac{\beta}{8} (u^2 + \upsilon^2)^2\;.  \eqno
(23)
$$
We regard Eq.~(23) as a Lagrangian for the coupled fields, $u$ and
$\upsilon$.  It implies that the field $u$ which varies the amplitude
of $\varphi$ has a positive mass (given by $\beta \sigma^2 = -\alpha =
|\alpha|$), while the variations of $v$ (phase) are massless.  These
massless excitations correspond to Goldstone bosons.

By analogy with the field theoretic considerations, one would
therefore expect existence of two modes of excitations in the broken
symmetry phase of systems described by a complex, non-conserved order
parameter such as the superfluid helium.  Two such modes -- second
sound phonons and rotons -- do indeed appear, but their correspondence
to the Goldstone bosons and massive excitations is at best imperfect.
We shall not pursue this aspect of the analogy further in our
discussion: In absence of a detailed theory of superfluid He$^4$ it is
hard to carry out such an investigation with a satisfactory degree of
rigor.

Let us now consider the case of a \underline{local} gauge theory.  The
corresponding Lagrangian is:
$$
L = -\frac{1}{4} B_{\mu\nu} B^{\mu \nu} + [(\partial_{\mu} + ie
A_{\mu}) \varphi^*] [(\partial_{\mu} - ie A_{\mu}) \varphi] - \alpha
\varphi^* \varphi - \frac{\beta}{2} (\varphi^* \varphi)^2 \ , \eqno(24)
$$
where $A_{\mu}$ is a massless gauge boson, and $B_{\mu \nu}=
\partial_{\mu} A_{\nu} - \partial_{\nu} A_{\mu}$.  In contrast to the
Lagrangian (21) which was invariant under global gauge
transformations, Eq.~(24) is invariant under the local Abelian
gauge transformation;
$$
U (\theta(x) ) = e^{-i\theta(x)}, \eqno(25)
$$
where $\varphi(x) \longrightarrow e^{-i\theta(x)} \varphi (x)$, and
$A_{\mu}(x) \longrightarrow A_{\mu} (x) - \frac{1}{e} \partial_{\mu}
\theta (x)$.

When we carry out an expansion around the local minimum of the
potential for the case $\alpha < 0$ we have considered previously, we
find:
$$
L = -\frac{1}{4} B_{\mu \nu} B^{\mu \nu}+ e^2 \sigma^2 A_{\mu} A^{\mu}
+ \frac{1}{2} (\partial_{\mu} u)^2 + \frac{1}{2} (\partial_{\mu}
\upsilon)^2 - \beta \sigma^2 u^2 - \sqrt{2}e \sigma A_{\mu}
\partial_{\mu} \upsilon + \ldots \eqno(26)
$$
The term involving $A_{\mu} A^{\mu}$ is the surprising outcome of the
symmetry breaking -- it looks as if the gauge vector field has
acquired a mass.

The Lagrangian (26) can be further simplified by fixing the gauge so
that $\theta (x)$ is equal to the phase of the original complex field
$\varphi (x)$.  In this gauge:
$$
\varphi (x) = \sigma + u(x) / \sqrt{2} \ ,  \eqno (27)
$$
and the Lagrangian becomes:
$$
L = -\frac{1}{4} B^{\prime}_{\mu \nu} B^{\prime \mu \nu} + e^2
\sigma^2 A^{\prime}_{\mu} A^{\prime \mu} + \frac{1}{2} (\partial_{\mu}
u)^2 - (\beta \sigma^2) u^2 - \frac{1}{8} \beta u^4 + \frac{1}{2} e^2
(A^{\prime}_{\mu})^2 (\sqrt{2} \sigma u + u^2)\;.  \eqno(28)
$$

In this form it is apparent that $L$ describes interaction of the
massive vector boson $A_{\mu}^{\prime}$ with a real scalar field $u$,
(the ``Higgs boson'') with the mass squared given by a $\beta \sigma^2
= -\alpha$.

\bigskip
\noindent{\bf Summary}\\
\indent The analogy between symmetry breaking in superfluid helium and in the
field theories with global gauge
invariance is now apparent: Wavefunction of the Bose condensate is the
analog of the scalar field.  Static global cosmic strings and vortex
lines in He II are a solution to an \underline{identical}
time-independent equation of a form:
$$
\bigtriangledown^2 \varphi \sim -\alpha \varphi + \beta |\varphi|^2 \varphi \ .
 \eqno(29)
$$
Their structure is defined by the correlation length $\xi$, which sets
the size of the core of the vortex. The total ``string tension'' is
associated with the kinetic energy of circulation, and is (in both
cases) logrithmically dependent on the cutoff at large scales.

Analogy in the case of the local gauge is even more striking.  There
the scalar field corresponds to the Bose condensate of Cooper pairs,
and the gauge field which acquires mass in the broken symmetry phase
is analogous to electromagnetic (gauge) field which becomes massive in
the superconducting state.  There are now two characteristic lengths
-- the correlation length $\xi$ of the scalar Bose condensate and the
penetration depth $\lambda$.  The analogy between the superconducting
vortex and the Nielsen-Olesen string was already pointed out.

\bigskip
\bigskip
\noindent {\bf 3.\quad FREEZE-OUT OF TOPOLOGICAL DEFECTS IN RAPID PHASE
TRANSITIONS}\\

In the preceding section, we have focused on the analogies between the
static properties of the broken symmetry phase.  In particular, we
have identified topologically stable time-independent solutions in
both superfluids and in superconductors.  Here we shall quantify the
general considerations of the first, introductory section, and derive
the density of vortex lines in superfluid He$^4$ which can be obtained
throughout the rapid (pressure) quench.  This experiment was
originally suggested$^{4,5,13}$ for the $\lambda$-line transition into
the superfluid He II, but its analogs can be carried out in other
phase transitions.  Slightly different (because of the first order
nature of the phase transition) case of liquid crystals was studied
experimentally by Yurke and his collaborators$^{6,7}$ and by Bowick
and his colleagues$^{10}$.

The case of He$^4$ has been implemented more recently by McClintock and his
colleagues$^{8,9}$.  It will be the focus of this section.  We shall review
the ``freeze-out'' scenario, argue that Ginzburg temperature does not play the
key role in determining the initial density of defects, estimate vortex line
density from the freeze-out argument, and show that it is consistent
with the experiment.  We shall also consider the case of
superconductors.

\bigskip
\noindent {\bf Quench into Superfluid}\\
\indent The transition from normal to superfluid He$^4$ is a particularly
suitable (from the
experimental point of view) analog of the cosmological phase
transitions.  In addition to the parallel between the Bose condensate
and scalar fields we have discussed in the last section, pressure
quench through the $\lambda$-line offers quick (dynamical) method of
reaching the state with the broken global gauge symmetry$^{4,5,13}$.
Moreover, the speed of the first sound (which will limit the rate at
which change of the pressure will be communicated through the medium)
is -- near the superfluid phase transition T$_{\lambda}$ -- orders of
magnitude in excess of the second sound velocity, which limits the
speed with which perturbations of the Bose condensate can spread.
Thus, we can reproduce in the superfluid phase transition the
``acausal'' nature of the cosmological phase transitions -- dynamics
of the broken symmetry phase is slow compared to the dynamics of the
quench.

Before the phase transition we imagine the order parameter which can
locally (i.e. within the $\xi$-sized region) fluctuate, but which is
on the average in the symmetric state.  This initial state will change
on the characteristic relaxation timescale:
$$
\tau = \hbar/ |\alpha| \ .  \eqno (30)
$$
This timescale will be essential in our considerations.  We shall use
it to estimate the expected density of vortex lines in the superfluid
through the following freezeout scenario: We note that as the pressure
drops, the dimensionless temperature parameter:
$$
\epsilon (T) = (T - T_{\lambda}) / T_{\lambda} \ ,  \eqno(31)
$$
which is positive in the normal He I, decreases and becomes negative.
When $T$ is close to $T_{\lambda}$ relaxation time becomes very large,
and it blows up to infinity on the $\lambda$-line;
$$
\tau = \tau_0 /|\epsilon|  \ . \eqno(32)
$$
Thus, the order parameter (which changes on the timescale given by
$\tau$) will be able to adjust only very slowly to the changes of
thermodynamic parameters.  This \underline{critical slowing down} is
accompanied by an increase of the correlation length $\xi$, which
diverges as:
$$
\xi = \xi_0 | \epsilon |^{-\nu}  \eqno(33)
$$
in the vicinity of $\epsilon = 0$.  Landau-Ginzburg theory with $\nu =
\frac{1}{2}$, and $\xi_0 = 5.6 \AA$ provides an acceptable (although
not an optimal, as we shall see shortly) fit for superfluid helium.

In course of a quench we shall imagine that -- very near the
$\lambda$-line -- dimensionless temperature will be approximately
proportional to the time before (after) $T = T_{\lambda}$ is reached:
$$
\epsilon = t / \tau_Q.   \eqno(34)
$$
Here $\tau_Q$ is the \underline{quench timescale}.

When $\tau \ll t$ the order parameter reacts to the quench-induced
change of $\epsilon$ by adjusting its average state (i. e., by
increasing the correlation length) in an essentially adiabatic
fashion: For each new value of $\epsilon$ there is a new realization
of a near-equilibrium with the approximately adjusted correlation
length, average size of $|\Psi|^2$, etc.  However, as $T$ approaches
$T_{\lambda}$, $\epsilon$ decreases, and the relaxation timescale
becomes larger and larger.

At some point the rate at which the thermodynamic changes occur will
become greater than the rate with which the order parameter $\Psi$ can
adjust.  At that instant our ``adiabatic'' approximation will cease to
be sufficient: Very close to $T_{\lambda}$ the configuration of the
order parameter will be essentially ``frozen out'' by the sluggishness
of its dynamics.  ``Impulse'' approximation will now apply: order
parameter will not be able to adjust -- the correlation length will
not be able to increase so as to keep up with its equilibrium value
prescribed by Eq.~(32).

The time $\hat{t}$ when the crossover from the ``adiabatic'' to the
``impulse'' regime occurs is of course critically important: It will
determine the size of the ``frozen out'' domains.  We can compute
$\hat{t}$ by equating the time-dependent relaxation time with the time
from the phase transition;
$$
\tau (\hat{t}) = \hat{t} \ ,  \eqno (35)
$$
or;
$$
\tau_0 / (\hat{t} / \tau_Q) = \hat{t}  \ , \eqno (36)
$$
which finally yields:
$$
\hat{t} = \sqrt{\tau_0 \tau_Q}  \ . \eqno (37)
$$
We are now nearly done.  To obtain the initial density of the
quench-generated vortex lines we still need to compute the
characteristic correlation length set by the ``freeze out'' at $t =
\hat{t}$ (or for $\epsilon = \epsilon ({\hat{t}})$).  This yields:
$$
d = \xi (\hat{t}) = \xi_0 (\tau_Q/ \tau_0)^{\nu/2} \ . \eqno (38)
$$
In the Landau-Ginzburg theory $\nu = \frac{1}{2}$.  The corresponding
vortex line density is then given by:
$$
\ell = k/d^2   \ ,\eqno (39)
$$
where $k$ is a proportionality constant which is thought to be smaller
then, but probably of the order of 1.

Equation (39) can be further rewritten to yield density of the vortex
lines in terms of the quench timescale:
$$
\ell = (k/\xi_0^2) \cdot (\tau_0/\tau_Q)^{\nu}   \ . \eqno (40)
$$
The exponent $\nu = \frac{1}{2}$ in the Landau-Ginzburg theory, but as
we have already mentioned, Landau-Ginzburg theory is only
qualitatively correct for superfluid helium.  Renormalization group
prediction for $\nu$ is;
$$
\nu = \frac{2}{3} \ .  \eqno (41)
$$
Experiments seem to agree with Eq.~(41).  Thus, for example the
correlation length determined from measurements is very well fit by:
$$
\xi(\epsilon) = \xi_0 / | \epsilon|^{0.67} \ ,   \eqno (42)
$$
where $\xi_0 = 4 \AA$.  Similarly, the velocity of the second sound is
approximately given by:
$$
u = u_0 |\epsilon|^{1-\nu} \ , \eqno (43)
$$
where $u_0 \cong 47$ m/s which is suitable for the renormalization
group exponent $\nu = \frac{2}{3}$ yields a better fit than the still
acceptable $u_0 \cong 70$ m/s, $\nu = \frac{1}{2}$ which obtains from
the Landau-Ginzburg theory.

\bigskip
\noindent {\bf Ginzburg Temperature and the Thermal Activation of Defects}\\
\indent For contrast, let us consider the activation mechanism for the
formation of topological defects.  The idea is quite
straightforward: For some range of temperatures below the phase
transition temperature $T_C = T_{\lambda}$ thermal fluctuation will be
able to ``flip'' the order parameter in the system between the local
minima around the rim of the ``Mexican hat'' potential.  Regions which
will undergo such transitions will have sizes of the order of the
correlation length $\xi$.  Therefore, the specific free energy barrier
computed from Eq.~(1) is:
$$
F(0) - F(\sigma) = \alpha^2 / (2\beta) \ .  \eqno (44)
$$
The energy of a volume of the size of a correlation length will be
comparable to the available thermal energy when:
$$
\xi^3 \alpha^2/2\beta \cong k_B T \ .  \eqno (45)
$$
Since $\xi \cong \hbar / \sqrt{ 2 m |\alpha|}$, the left hand side of the
above equation varies as $(T-T_C)^{1/2}$.  A numerical estimate for
the Ginzburg temperature $T_G$ for which Eq.~(45) is satisfied yields;
$$
T_\lambda - T_G \cong 0.5\; [^{\circ} K] \ , \eqno (46)
$$
for superfluid Helium II.  This estimate is in reasonable agreement
with the width of the so-called $\lambda$ anomaly in the specific heat
(the shape of which is responsible for the name ``$\lambda$-line'' of
the transition from normal to super phase of liquid helium), but it
seems to have little bearing on the existence or stability of vortex
lines.  We shall confirm this immediately below, while comparing
theoretical prediction of vortex line production in a quench with
experimental results.  Indeed, this conclusion seems sensible and
valid for superfluids even without appealing to the recent
experiments: For, if small scale (correlation length $\xi$)
fluctuations between the degenerate minima of the potential could
create topological defects, then this process would also destroy them
on the timescale given by $\tau$.  This is known not to happen to
vortex line density generated by various methods in a superfluid above
the Ginzburg temperature $T_G$ (but, of course, below $T_{\lambda}$).

The reason $T_G$ does not play a decisive role in creation of vortex
lines in He$^4$ (and, for that matter, should not be an important
mechanism for production of other topological defects in quench -
induced phase transitions either in condensed matter or in cosmology)
is likely to be associated with the spatial
extent of the thermally activated transitions: Local thermal
fluctuation can perhaps create small loops of vortex line, but these
loops will have a radius approximately equal to the size of their core
(since both are defined by the same correlation length $\xi$).  Such
ill-defined vortex line ``doughnuts'' are unlikely to be stable --
after all, they represent a configurations which may be a shallow
local minimum of the free energy, but which have a higher free energy
then the uniform superfluid.  Hence, they are not large enough to be
really topologically stable -- change of the field configuration in a
finite region of space the order of $\xi$ suffices to return to the
uniform ``true vacuum''.  (Indeed, one is tempted to speculate that
the rotons -- which become plentiful above $T_G$ -- are thermally
excited in such fashion).

It is quite evident that such local loops cannot result in creation
(or destruction) of one long vortex line which is likely to be the
dominant contribution to the vortex line density following pressure
quench. Long lines or large loops created by the freezeout can become
``wrinkled'' as a result of thermal activation on the Ginzburg scale,
but this is not expected to lead to a significant increase of $\ell$,
Eq. (40).  Similar reasoning can be repeated for membranes.  Such
requirements are least convincing in the case of monopoles (which have
spatial extent of order $\xi$ in every direction).  However even
monopoles will have to be created ``in pairs'' by thermal
fluctuations.  These pairs of monopoles of opposite charge will be
separated by distance of order $\xi$.  Therefore, they are not likely
to separate and survive.

We are led to the conclusion that the dominant process in creation of
topological defects will have to do with the critical slowing down and
a consequent ``freeze-out'' of the fluctuations of the order parameter
at the time $\hat{t}$ rather than with the thermal activation and
Ginzburg temperature.  We shall evaluate this conclusion in the light
of experimental results below.

\bigskip
\noindent
{\bf Comparison with the Experiment}\\ \indent The experiment carried
out at the Lancaster University$^{8,9}$ follows the pressure quench
strategy proposed for the superfluid helium one decade ago.$^{4,5,13}$  The
typical $\Delta \epsilon$ -- change of the relative temperature -- can
be crudely estimated from the initial and final temperature
differences:
$$
\Delta \epsilon \cdot T_{\lambda} \  \simeq \  [T_i - T_{\lambda} (p_i)] \ -\
[T_f -
T_{\lambda} (p_f)] \ , \eqno (47)
$$
where $T_i$, $p_i$ and $T_f$, $p_f$ are the initial and final
temperature and pressure respectively.  This method yields $\Delta
\epsilon \sim 0.1$, which, combined with the time interval $\Delta t
\sim 3$ ms over which pressure drop occurs$^8$, results in the
estimate of the quench timescale:
$$
\tau_Q \simeq \Delta t / \Delta \epsilon \simeq 30 {\rm ms}.  \eqno (48)
$$
I would like to emphasize that this is a very rough estimate of the
actual $\tau_Q$.  To obtain a more reliable $\tau_Q$ one would have to
compute quench rate along the isentrope ($S=$const) of the quench:
$$
(\tau_Q)^{-1} = (\partial \epsilon (t) /\partial t)_{S = {\rm const},
\epsilon = 0} \ , \eqno (49)
$$
for the point at which the phase transition occurs (that is, for
$\epsilon = 0$).  I have not carried out such a calculation, but the
shapes of the isentropes near $T_{\lambda}$ lead one to believe that
$\tau_Q$ given by Eq.~(49) would be somewhat (perhaps even by an order
of magnitude or so) larger than the simpler estimate of Eq.~(48) would
have it.

With these estimates (and the associated caveats) in mind let us now
use Eq.  (40) to estimate initial vortex line densities.  For the
Landau-Ginzburg theory ($\nu = \frac{1}{2}$ with $\xi_0 = 5.6 \AA$ and
$\tau_0 = 0.85 \cdot 10^{-11}$ s) one obtains a prediction:
$$
\ell_{LG} \cong 3 ~ ( \tau_Q/100~ {\rm ms})^{-1/2} \cdot 10^{13}  \rm m^{-2} \
, \eqno (50a)
$$
while for renormalization group theory the prediction is:
$$
\ell_{RG} \cong  1.2 ~ (\tau_Q/100~{\rm ms})^{-2/3} \cdot 10^{12} \rm m^{-2} \
.  \eqno (50b)
$$
These predictions bracket the lower bound of $\ell \sim 10^{13} \rm
m^{-2}$ based on the experimental results.  Vortex line production
occurs also where a quench is initiated just below ($\sim 10$~mK) the
$\lambda$-line. There is, however, no noticeable vorticity produced in
quenches which start far below the phase transition.  This is an
intriguing observation.  A possible explanation of this effect is to
appeal to the combination of the flows which are also induced by the
quench$^9$ and the thermally activated vortex line production.  What
can happen is that rare, thermally excited vortex lines get stretched
and entangled by the flows generated in course of the expansion of the
He$^4$ container.$^9$ This may result in a turbulent tangle of vortex
lines, amplifying the pre-existing thermally activated ``seed''
vorticity.  If this mechanism does indeed operate, it is likely that
flows also amplify the density of vortex lines in quenches which cross
$T_{\lambda}$.

There are also serious concerns about the accuracy of the estimate
derived from the ``naive'' application of the freezeout scenario to
superfluid He$^4$.  Thus, for example, the estimated ``freeze out''
correlation length (Eq.~(38)) is only an order of magnitude estimate
of $d$.  Indeed, it seems likely that the actual correlation length
will be bounded from above by the freeze out scale, but could be
somewhat (by a factor of a few) smaller than $\xi (\hat t)$, Eq.~(38).
Moreover, an accurate estimate of $k$ of Eq.~(40) would be useful.
Last but not least, the very process of formation of the Bose
condensate is unlikely to be instantaneous, and -- as it was already
pointed out$^{13}$ -- large vortex line densities may depress
$T_{\lambda}$.

In spite of these reservations, the obvious conclusion of this section
is that the rapid quench generates vortex line density
consistent with the theoretical predictions based on the idea of the
``freeze out'' of the configurations of the order parameter.$^{4,5,13}$
Moreover, it is also clear that the Ginzburg temperature does not play
as decisive a role as it was originally expected$^2$.

\bigskip
\noindent
{\bf Quench into the Superconducting State}\\ \indent Much of what we
can anticipate in the case of a quench into a superconductor will be
based on an argument which parallels the case of superfluid He$^4$.
We shall therefore be brief in the discussion of vortex line creation
in (type II) superconductors.  In effect, we shall repeat what was
already said before, but we shall also emphasize the differences
between the two cases.

It should be noted that some of these differences are non-trivial.
Thus, presence of the gauge field in superconductors complicates the
problem by adding extra physics which leads, for example, to an
additional characteristic scale.  On the other hand, superconductors
are rather well described by the simple Landau-Ginzburg theory with
the order parameter representing the wavefunction of the Bose
condensate of Cooper pairs.$^{12}$

Let us begin by noting that, in contrast to just one superfluid
He$^4$, the list of superconducting materials is very long (even if we
restrict ourselves to type II only).  This is especially true if one
includes in it new ``high $T_C$'' superconductors.  In general,
temperature of the phase transition into the superconducting state is
\underline{not} very sensitive to pressure.$^{14}$ Nevertheless, some
sensitivity to pressure does exist (especially in the high-$T_C$
materials) so one could contemplate a similar ``quench'' scenario as
the one described in the case of He$^4$.  This would have the
advantage of achieving the transition very quickly and in a manner
which does not directly involve electrons.  However, if the pressure
quench proved to be impossible, cooling might be an interesting
alternative, and could be achieved relatively rapidly when the sample
is a thin layer of superconducting material.

This 2-D (rather than truly 3-D) strategy may have one more advantage:
Vortex lines which form inside the superconductor cannot be really
``seen'' from the outside (in superfluid helium second sound
attenuation can be used to measure vortex line density).  And since
superconducting vortices in the bulk are invisible, one may as well
concentrate on the points where they enter or leave the sample, which
suggests very flattened samples.

Two-dimensional geometry may also help address one more likely
problem: Vortex lines created in the superconductor will have a
tendency to annihilate or to escape from the sample (after all, the
state without topological defects and without the associated magnetic
fields has lower energy!).  Using flattened samples may allow one to
trap (pin) vortex lines (and thus slow down annihilation).

With all these caveats in mind, let us now estimate the density of
vortex line generated in a rapid quench in the type II superconductor.
Proceeding along the path parallel to the one we have adopted for the
superfluid phase transition, we are led to evaluate the relaxation
timescale of the order parameter $\tau$ for superconductors; $ \tau =
\tau_0 / |\epsilon| $ (see Eq.~(32)) where $\tau_0$ can be
approximately computed from the so-called Gorkov equation to be:
$$
\tau_0 = \frac {\pi \hbar } { 16 k_B T_C }  \ . \eqno (51)
$$
The freeze out time will be still given by $\hat{t} = \sqrt{\tau_0
\tau_Q}$, Eq.~(37), which with the help of Eq.~(51) can be numerically
evaluated to be:
$$
\hat{t} \cong 1.225 \cdot \sqrt{\tau_Q /T_C}\;  [\mu s]\;.  \eqno (52)
$$
when the quench timescale $\tau_Q$ is in seconds and $T_C$ in degrees
Kelvin.  There is of course no guarantee that the approximations which
lead to Eq.~(51) will be accurate for all of the superconducting
materials, but the above estimate of $\hat{t}$ is likely to give an
order of magnitude value for the freezeout time.

The size of the frozen-out domains will be given by Equation (33) with
the exponent $\nu = 1/2$:
$$
\xi = \xi_0 / \sqrt{|\epsilon|} \ .  \eqno (53)
$$
Typical values of $\xi_0$ in superconductors are significantly larger
than in superfluids (i.e. $\xi \sim 1000 \AA = 10^{-5}$ cm) although
smaller $\xi_0$ can also be found (for example, in high-temperature
superconductors).  Nevertheless, for the purpose of rough estimates
one can evaluate $d$ to be:
$$
d \simeq 10^{-2} (\xi_0 / 1000 \AA) \tau_Q^{1/4} ~ [{\rm cm}] \
. \eqno (54)
$$
Hence, the domain sizes are much larger in superconductors than in
superfluids.  The resulting vortex line density will be therefore
correspondingly smaller:
$$
\ell \cong 10^4 (1000\AA/\xi_0)^2 / \sqrt{\tau_Q} ~ {\rm cm}^{-2} \cong 10^8
 (1000\AA/\xi_0)^2 / \sqrt{\tau_Q} ~ {\rm m}^{-2}.
\eqno (55)
$$
These estimates may still look reasonably hopeful, but it seems
unlikely to this author that the values of $\ell$ predicted by
Eq.~(55) will be easily detectable experimentally: As is the case in
the superfluid helium (or in a liquid crystal) domain structure will
give rise to the initial network of the flux lines with $\ell$
predicted above, but the evolution -- shrinking of the loops,
straightening of the long string -- will quickly lower the value of
$\ell$.  One could slow down this process by choosing a material with
plentiful pinning sites.  However, the presence of the inhomogeneities
which give rise to pinning may also invalidate some of the arguments
we have put forward above by -- for example -- making the phase
transition temperature $T_C$ location-dependent.  Indeed, this last
remark emphasizes one of the great advantages of superfluid helium
from the point of view of quench experiments$^{8,9}$ -- its
homogeneity.

In view of these considerations, it is clear that the experimental
study of flux line creation in rapid phase transitions into the
superconductor is bound to be more complicated than in the case of
superfluids.  These difficulties are mainly of the experimental
nature.  In particular, both rapid quench and the detection of flux
lines appear to be much more difficult to accomplish in
superconductors than in superfluid helium.  Both of these difficulties
may be partially alleviated by using two-dimensional samples with the
``thin'' dimension somewhat larger than $d$, Eq.~(54), but much
smaller than the other two dimensions.  This geometry could help in
cooling, and would also allow easier access to the potentially
detectable ``ends'' of flux tubes.

\bigskip
\bigskip
\noindent
{\bf 4.\quad QUENCH IN AN ANNULUS}\\

Complexity of evolution of the Brownian network of vortex lines makes
it worthwhile to discuss a conceptually simpler version of the rapid
phase transition into a vacuum with a non-trivial topology of the
ground state manifold.  With this motivation in mind, we shall
consider creation of Bose condensate in superfluid He$^4$ in an
annulus.  Similar experiments can be also performed in superconducting
loops.  In the superfluid helium, I shall argue, rapid quench will set
up a deficit of the phase $\theta$ which distinguishes between the
various degenerate vacua in the broken symmetry state.  This phase
difference will result in a flow in a random direction, but -- for
reasonable quench timescales -- with a detectable velocity ($\sim
0.1$~cm/s).

Quench into a superconductor will similarly lead to a phase deficit.
This phase deficit will translate into a supercurrent, which will trap
flux quanta inside the loop.  The number of the trapped quanta will,
in general, depend on the rate of the phase transition, but may be
also influenced by other factors which are normally disregarded in the
discussions carried out in the cosmological context such as the
inductance and resistivity of the L-R circuit equivalent to the loop.
Moreover, geometry of the loop can be made (approximately)
one-dimensional, which may (in the appropriate conditions) restore the
importance of activation processes.  Thus, in one-dimensional
superconductors the original version of the Kibble mechanism with its
emphasis on thermal activation may be again important (although for
reasons which are not expected to be valid in the cosmological
context).

\bigskip
\noindent
{\bf Phase Around the Loop: Generating Persistent Superflows With a
Quench}\\ \indent Let us consider an imaginary circular loop of radius
$r$ in a bulk superfluid.  After the quench the circumference of this
loop will intersect approximately:
$$
N = 2\pi r / \xi (\hat{t}) = 2 \pi r / d \eqno (56)
$$
independent domains.  Hence, the anticipated phase mismatch will be of
the order of;
$$
\Delta \theta \sim \sqrt{N} = \sqrt {\frac{2\pi}{d}} ~ .  \eqno (57)
$$
Consequently, the gradient of the average phase will be approximately
equal to:
$$
g = \Delta \theta/2\pi r = \sqrt{1 / 2\pi r d} \ .  \eqno (58)
$$
In the superfluid, such gradient of the phase implies supercurrent
velocity of:
$$
v = (\hbar / m) g = (\hbar / m) / (C d)^{1/2} \ , \eqno ( 59)
$$
where $C = 2\pi r$ is the circumference of the loop.

It was already pointed out some time ago$^{4,5}$ that this phase
difference (which would decrease in bulk superfluid with the evolution
of the vortex line network) can be ``frozen out'' by performing the
quench in an annulus.  Moreover, the resulting velocities are
measurable ($\sim$ mm/s) and depend only weakly on the quench
timescale;
$$
v \sim \tau_Q^{-\frac{\nu}{4}} \approx \tau_Q^{ - \frac 1 6} \ , \eqno
(60)
$$
where we have used the renormalization group value of $\nu = 2/3$.
The corresponding angular momentum is non-negligible, but there is no
paradox, as its origins can be traced to Brownian motion at the
``freeze-out time'' $\hat{t}$.$^5$

The principal advantage of the quench in annular geometry is the
time-independence of the effect -- persistent supercurrent.  In
contrast to the vortices created in the bulk superfluid, persistent
supercurrents do not decay, or at least do not decay on a rapid
timescale on which vortex lines intersect and disappear.  Hence, one
may have a better chance to obtain an estimate of the frozen-out
correlation length.  On the other hand, this experiment is
significantly more challenging than the bulk version.  Among the
experimental difficulties one should list the problem of performing
the quench in an axially-symmetric fashion (so that the superfluid is
not ``pushed'' in the process) as well as the measurement of the
resulting velocity.

There are also theoretical complications: Rapid quench will change the
equilibrium correlation length $\xi (T)$ from a large near - $T_C$
value to much smaller value far from the $\lambda$-line.  This
changing value of $\xi (T)$ will be, at some stage, comparable to the
small radius $a$ of the torus containing the superfluid.  When $a <
\xi (T)$, the superfluid is effectively one-dimensional.  Thus, vortex
lines cannot ``fit'' within the annular container.  Moreover,
activation energy required to change the winding number $n_w$ defined
as;
$$
n_w = \frac {\Delta \theta } { 2\pi } \eqno (61)
$$
is given by;
$$
\Delta F_0 \cong \pi a^2 \xi (T) \Delta F \ .  \eqno (62)
$$
When $a < \xi, \Delta F_0$ is \underline{less} than the energy
one would normally employ in deriving Ginzburg temperature.  For this
reason, in the first papers on this subject$^{4,5}$ I have suggested
using annulus with $\xi (\hat{t}) \sim a$: For $t > \hat{t}\; (\xi (t)
< \xi (\hat{t}))$ coherent fluctuations of the volume $\sim a^3$ will
become unlikely, so that the winding number will be ``safe'' from
thermally activated processes.  This conclusion -- while essentially
valid -- ignores creation of small sections of the vortex line inside
the annulus in the regime where $a \sim \xi (t)$.  Such vortex lines
may change the average velocity of the superflow.  With time, they may
also migrate towards the inner (or outer) wall of the annulus, thus
changing the winding number.

In spite of the above concerns, I believe that this ``phase around the
loop'' experiment is very much worth performing: It offers a dramatic
demonstration of the ``phase freezeout'' predicting generation of a
significant velocity (and of a measurable angular momentum) as a
result of the rapid phase transition.  Moreover, the possibility of
thermally activated transitions should be regarded not just as a
complication, but as an opportunity.  Thus, for example, one can
contemplate studying of not just quench - generated superflows, but
also decay -- due to thermally activated processes -- of the winding
number.  In this regime one is probing the interplay between thermal
activation and topological stability.

\bigskip
\noindent
{\bf Winding Number in a Superconducting Loop: Quenching out Flux}\\
\indent For the quench experiments carried out in a superconducting
loop the basic scenario of locking out superflow (of Cooper pairs)
should be still applicable, although with a few important (and
interesting) complications.  Let us therefore consider a loop of some
radius $r$ with the wire diameter given by $2a$, where $a \ll r$.  We
shall suppose that for the quench timescale under consideration the
frozen-out correlation length $d$ is at most of the order of $2a$, so
that one has typically no more than one domain across the wire.  Rapid
transition will then result in a typical phase difference $\Delta
\theta$ along the circumference of the wire in accord with Eq.  (57).
Hence, the resulting winding number $n_{\Phi}$ (and the number of
trapped flux quanta) should be of the order:
$$
n_{\Phi} = \frac { \Delta \theta} { 2\pi } = (2\pi)^{-1} \sqrt{ \frac
C d} \ .  \eqno (63)
$$
For a loop of $r = 1$~cm $(C=2\pi r)$ and the frozen out correlation
length of $\sim 10^{-2}$ cm (see Eq.~(54)) this yields small but
easily measurable $n_{\Phi} \sim 3$.

So far we have ignored the role of the gauge field: Our prediction,
Eq.  (63), is based solely on the fate of the order parameter.  Yet,
the energy of the trapped flux $E_{\Phi}$ can easily be comparable or
even larger than the energy of thermal fluctuations at the temperature
at which the phase transition is taking place.  Thus:
$$
E_{\Phi} = \Phi^2 / 2L = \frac{ n^2_{\Phi} (h c/ 2e)^2} { 2 L } =
n^2_{\Phi} E_0 \ , \eqno (64)
$$
where $L$ is the self-inductance of the loop. It is given approximately by:
$$
L \cong 4\pi \cdot 10^{-9} ~ r ~ \ln (r/a) [\rm H] \ , \eqno (65)
$$
where $L$ is measured in Henry's and $r$ in centimeters.  Energy unit
$E_0$ in Eq.~(64) stands for the energy associated with a single
quantum of flux trapped in a loop of self-inductance $L$;
$$
E_0 = \Phi_0 ^2 /(2 L) \simeq 2.5 \cdot 10^{-16} r^{-1} [\rm erg] \ ,
\eqno (66)
$$
where we have taken $r/a = 1000$.  For comparison, the energy of
thermal fluctuations available at the critical temperature is:
$$
E_T = \frac{1}{2} k_B T_C \cong 7 \times 10^{-17} \cdot T_C ~[\rm erg]
\ . \eqno (67)
$$
where $T_C$ is given in degrees Kelvin.  For typical superconductors
$T_C$ falls in the range of 0.1 to few tens of Kelvins.  Thus, even in
the relatively high-temperature cases, only a few quanta of the frozen
out flux could come directly from thermal fluctuations.  Moreover, the
value of the inductance (and, consequently, $E_0$) can be altered by
changing the geometry of the loop.

We have therefore three possible sources of the locked-out flux.  The
first (trivial) is the background flux which is going to determine the
average flux through the loop.  Dispersion about that average can be
either due to the freeze-out of thermal fluctuations of the flux, in
which case;
$$
\frac{\delta \Phi^2_T}{2L} = \frac{1}{2} k_B T_C ~ ,  \eqno (68)
$$
or it can have the more interesting origin in the freeze-out of the
order parameter we have considered above, in which case the dispersion
of the flux will be given by Eq.~(63), or:
$$
\delta \Phi^2 = \Phi^2_0 \cdot n_{\Phi}^2 = ( \Phi^2_0 / (2\pi)^2) (C/d) \ .
\eqno (69)
$$

It should be (at least in principle) relatively easy to distinguish
between these two cases.  Freeze-out of the order parameter results in
the prediction which is independent of the inductance $L$ of the
superconducting loop, but which depends on its circumference $C$.
Thus, one can tell the difference between the two cases by deforming
the loop (i.e., by coiling it up) which will change $L$, but leave $C$
unaltered.  Moreover, freeze-out of the order parameter should result
in slow variations of the flux:
$$
n_{\Phi} \sim \tau_Q^{-1/8} \ .  \eqno (70)
$$
with the quench timescale. In addition to those signatures one can of
course compare the absolute value of the predicted flux variations.

It should be emphasized that the question which is being posed here is
a profound one: We are trying to determine whether it is the gauge
field or the order parameter which controls the final frozen-out
winding number.  In the discussions carried out in the cosmological
context the usual assumption associates topological defect formation
with the correlation length of the order parameter.  If this were
indeed the case, then the estimate given by Eq.~(69) should prevail.
There are however circumstances in which thermal fluctuations of the
gauge field (rather than of the order parameter) may emerge from the
experiments.  The first one is relatively trivial: Let us suppose that
the phase transition does not happen simultaneously around the loop,
but there is one spot which becomes superconducting last.  Then, for a
while, the order parameter along the superconducting section of the
loop (and the flux associated with it) will be in contact with a
thermal reservoir at (approximately) $T_C$.  The whole of the loop is
\underline{not} superconducting.  Therefore, the flux inside it is
still able to vary, presumably on a timescale associated with the
inductance:
$$
\tau_{RL} = L/R  \ , \eqno (71)
$$
where $R$ is the resistance of the normal section of the loop.
Consequently, as the still normal section of the loop becomes
superconducting (thus fixing the value of the flux) the order
parameter coupled with the flux of the magnetic field continues to be
driven by thermal fluctuations.

In this limit, quench becomes reminiscent of the experiment carried
out by Tate {\it et~al.}$^{15}$ They have repeatedly heated a small
section of superconducting niobium ring.  Upon cooling, the ring had
contained a random number of quanta with a dispersion corresponding to
6.78 $^{\circ}$K which is somewhat less than the critical temperature
of pure niobium (which is 9.17 $^{\circ}$K).  It is clear that in this
limit (when the loop is kept open for periods of time long compared to
all other physically relevant timescale) the number of quanta
eventually locked out will be determined by thermal fluctuations.

It should be noted that the geometry of the loop which is undergoing
the quench is different from the one relevant in the cosmological
context. The key consequence of this difference has to do with the
fact that inside the loop gauge field remains massless even after the
phase transition. This is clearly not going to be so in the early
universe, where the critical temperature is reached more or less
simultaneously throughout the volume. Nevertheless, experiments
proposed here should allow one to shed a new light on the relative
roles of the gauge field and of the order parameter in general. This
point is of significant interest, and related questions about the
cosmological phase transitions have also been raised.$^{16}$

The crucial point which is relevant for the loop geometry in this
context concerns the nature and the rates of the processes which could
alter the winding number between the instant of its creation $\hat t$
and the later time when it can be measured. Two of them can be readily
identified: (i) Ginzburg - like activation which acts locally (that
is, within the correlation length $\xi$) and on the order parameter,
and (ii) Transitions involving the whole loop and mediated by the
(still massless) gauge field trapped inside.  Both of these processes
will be effective only near the critical temperature.  Thus, once the
relatively narrow (few milikelvin, at the most) danger zone is
traversed in course of the quench, the number of flux quanta shall be
fixed and can be measured at leisure.

The role of the activation processes in thin wires (that is, in the
case when the radius of the wire is smaller then the correlation
length) has been studied (see chapter 7 of Ref. 12). It is determined
by the free energy increment needed to ``flip'' the order parameter
over the potential barrier in a small section ($\sim \xi$) of the
loop. The exact expression turns out to be given by:
$$ \Delta F_0 = \frac {4 \sqrt 2} {3} \frac {\alpha^2} {\beta} A \xi \
.  \eqno (72) $$ Above, $A$ is the cross section of the wire.

Theory and experiment both agree that unless $A$ is very small ($A
\sim 10^{-9} ~ \rm cm^2$) so that the volume of the section of the
wire over which the transition occurs is much less than $\xi^3$, such
transitions cannot be expected to take place within less than $\sim
10^{-3} ~^{\circ}$K of the critical temperature. As the temperature -
dependent terms in $\Delta F_0$ change with $|\epsilon|^{3/2}$, it
should be possible to select parameters so as to assure that the
Ginzburg - like process considered here will not change the winding
number. For example, the rate can be lowered by increasing the cross
section of the wire to $A \sim 10^{-4}~$cm$^2$ ($a \sim
10^{-2}~$cm). In effect, as was the case for the quench in the
``loop'' of the superfluid helium, one would want to select $a$ of the
order of $d = \xi (\hat t)$ to keep the problem effectively one -
dimensional, while, at the same time, making activation energetically
as expensive as in the bulk.

The second process which can influence the number of captured flux
quanta has a global nature (or, at the very least, its rate does not
depend on $\xi$) and is mediated by the magnetic field trapped inside
the loop. The activation energy associated with it can be estimated by
computing the kinetic energy of the Cooper pairs in the case when the
enclosed flux does not correspond to an integer number of flux quanta.

Just below the phase transition London penetration depth $\lambda$ can
be large in comparison with the diameter of the wire, $\lambda >
2a$. Then the current density in the wire will be (approximately)
independent of location (that is, same on the inside and on the
outside of the loop). The total energy (corresponding to a certain
current density) will comprise of two contributions: The energy of the
magnetic flux (given by Eq. (64)) and the kinetic energy of the charge
carriers. For an arbitrary flux one may choose $n_{\Phi}$ so that the
residual wave vector:
$$ q = (2 \pi / C) (n_{\Phi} - \Phi / \Phi _0) \ , \eqno (73) $$ is in
the interval $-\pi < q C < \pi$. This corresponds to the velocity:
$$ v = \hbar q / m_C \ , \eqno (74) $$ where $m_C$ is mass of the
Cooper pair. The total energy will be then given by:
$$ E = \Phi ^2 / (2 L) + N_C m_C v^2 / 2 \ , \eqno (75) $$ where $N_C$
is the total number of the Cooper pairs in the loop. $N_C$ can be (in
our case) obtained form the density of the pairs $n_C$, (which is
given by the square of the order parameter):
$$ N_C = C \; \pi a^2 \; n_C \ . \eqno (76) $$ Quantization of the
flux is the consequence of the existence of the local minima of $E$ as
a function of $\Phi$. Using Eq. (73) - (76) it is straightforward to
show that $E$ is minimized when;
$$ \Phi = n_{\Phi} \Phi_0 / ( 1 + E_0 / E_K) \ , \eqno (78) $$ where
$E_0$ corresponds to the energy of a single quantum of flux, Eq. (66),
and $E_K$ is the kinetic energy due to the velocity resulting from the
phase difference of $2\pi$ over the circumference $C$ and is given by:
$$ E_K = \frac {1} {2} N_C m_C [(\hbar/m_C) (2\pi/C)]^2 \ . \eqno (79)
$$

Consequently, in equilibrium the flux in a loop of a given
self-inductance $L$ and circumference $C$ will be quantized in the
units of$^{17-19}$:
$$ \tilde \Phi_0 = \Phi_0 / ( 1 + E_0 / E_K) \ . \eqno (80) $$
Moreover, quantization is not ``absolute'': Transitions between the
different minima of the total energy -- corresponding to different
integer values of flux -- are guarded against by the potential barrier
with the height:
$$ E_B \cong E_K/2 \ . \eqno (81)$$ At $t = \hat t$ this energy is
generally smaller than the free energy $\Delta F_0$, Eq. (72), and may
be of the order of $k_B T_C$. However, changes of flux cannot take
place on a timescale smaller than $\tau_{RL}$, Eq. (71). Moreover,
near $T_C$ the resistance $R$ of the loop is expected to rapidly
decrease, and $\tau_{RL}$ will increase.

It is difficult to tell whether these processes will allow the flux to
acquire its order parameter mandated value, Eq. (63), or if the
trapped gauge field will assume value closer to the thermal
prediction, Eq. (68). With Pablo Laguna, we are presently carrying out
a numerical study of rapid quenches in (quasi-) one-dimensional loop
configurations to clarify some of the relevant issues$^{20}$. The
above concerns notwithstanding, it may turn out that the most
challenging difficulties in carrying out actual experiments may be of
more ``mundane'' nature, and may involve making sure that the critical
temperature is reached (nearly) simultaneously around the loop, that
the loop is screened from outside magnetic fields, etc.

One could also contemplate artificially controlling the number of the
independent pieces of the superconducting order parameter (instead of
relying on $\xi$).  This could be achieved, for example, by heating up
the loop in many places, so as to break up the superconducting loop
into many pieces, (say, $N$), which should result in a phase
difference $\sim \sqrt N$ (in addition to the fluxoid due to the
thermal activation of the field, Eq. (68)), and a corresponding
winding number after the reconnection. Many variants of such
experiments are possible. We shall not analyze them here further.

\bigskip
\bigskip
\noindent
{\bf 5.\quad COSMOLOGICAL IMPLICATIONS}\\

Throughout much of this paper we have focused on condensed matter
systems, or, to be more specific, on condensed matter analogues of the
cosmological ``quench''. The central lesson of our discussion was a
revision of the original mechanism for defect formation in the course
of second order phase transitions involving a non-conserved order
parameter$^2$. We have concluded that the initial density of the
topological defects will be set by the correlation length at the
freezeout instant $\hat t$ -- that is, at the moment when the
relaxation timescale of the order parameter will be comparable to the
time from when the critical temperature $T_C$ is attained.$^{4,5,13}$
At that instant, as a consequence of critical slowing down,
perturbations of the order parameter become so sluggish that they
effectively cease to evolve, so that in the time interval ($-\hat t, \
\hat t$) the order parameter cannot adjust its correlation length
to the values of the thermodynamic parameters which are changing
as a result of the quench. When, below $T_C$, and for $t > \hat t$,
the dynamics will ``restart'', it will be too late to get rid of the
defects, which have been by then ``set in concrete'' (or, rather, in
the topologically stable configurations of the order parameter).

Experimental results obtained in the superfluid experiment by the
Lancaster group$^{8,9}$ demonstrate that this argument is unaffected
by the thermal activation process invoked in the original discussion
by Kibble$^2$.  Ginzburg temperature -- that is, the temperature at
which vacuum can be thermally ``flipped'' over the potential barrier
in regions of size $\xi$ -- is far below the critical temperature
$T_{\lambda}$ in the superfluid helium case ($T_{\lambda} - T_G \simeq
0.5 ~^\circ$K). In spite of that, there is no evidence for copious
thermal creation of vortex lines. We expect that the same situation
will prevail in superconductors, where $T_G$ is very close to $T_C$,
even though the freezeout temperature will be now typically lower than
$T_G$. This is not unexpected -- topological defects created by the
activation process are very small ($\sim \xi$). Therefore, topology
does not yet stabilize them: Decay of a doughnut - shaped string loop
can take place on a relaxation timescale, as it involves getting rid
of the ``hole'' in the center of the doughnut, and is obviously
energetically favored.

In superconductors, where $T_G$ can be expected to be above the
freezeout temperature reached during the quench at $\hat t$, that is,
within the ``time out'' interval ($ - \hat t, \hat t$), this
expectation is even easier to justify: The order parameter is simply
too sluggish to do anything between $T_C$ and $T_G$.  However, for
superconductors a different timescale and a different lengthscale are
also relevant. In addition to the order parameter (which is still
expected to play a pivotal role there is also the gauge field, which
may be quite important in the loop geometry, but may have some
significance also in the bulk$^{16}$. It is hoped that the experiments
suggested above can be carried out and that they will shed as much
light on the dynamics of phase transitions with the local gauge as the
Lancaster superfluid experiment already did for the case of the global
gauge$^{8,9}$.

Last but not least, let us briefly consider implications of our
observations concerning the relative roles of the freezeout instant
and Ginzburg temperature in the cosmological context. To carry out
this discussion, we shall adopt the usual high energy / cosmology unit
system with $\hbar=c=k_B=1$, and focus on the order parameter of the
effective field theory described by the Lagrangian given either by
Eq. (21) or (24). The principal difference with superfluids and
superconductors is due to the fact that the corresponding equation of
motion has a \underline{second} time derivative on the left hand side
(in contrast to the first derivative relevant for the
``nonrelativistic'' condensed matter cases). Hence, the relaxation
timescale of the order parameter will be given by:
$$ \tau = 1 / \sqrt {\alpha} = \tau_0 / \sqrt {| \epsilon |} \ . \eqno
(82) $$ That is, critical slowing down sets in with the inverse of the
\underline{square root} of the relative temperature, rather than with
$1/| \epsilon |$, as was the case for Eq. (32). One immediate
consequence of this difference is that the corresponding
characteristic velocity defined by $\xi / \tau$ with which
perturbations of the order parameter can spread will be finite and
relatively unchanged in the vicinity of $T_C$. By contrast, the second
sound velocity in the superfluid decreases as $|\epsilon|^{1-\nu }$ in
the vicinity of $T_C$.

The cosmological version of Eq. (35) for the freezeout instant can be
now written:
$$ \tau_0 /\sqrt{|\epsilon (\hat t)|} = \hat t \ , \eqno (83) $$
which, with the usual definition of the quench timescale ($\epsilon =
t / \tau _Q$, Eq. (34)) yields:
$$ \hat t = \tau_0^{\frac 2 3} \tau_Q^{\frac 1 3} \ . \eqno (84) $$
This is in contrast to the previously derived and more symmetric $\hat
t = \sqrt {\tau_0 \tau_Q}$, Eq. (37), which is valid for the
superfluid and for the superconductor case, when $\tau \sim 1/|
\epsilon|$.

We are now ready to compare and contrast prediction for the freezeout
scenario with the Ginzburg activation mechanism. In the radiation -
dominated era the temperature $T$ and the time $t$ (which is measured
form the beginning of time -- from the Big Bang) are related with a
simple equality:
$$ T^2 t = \Gamma M_{Pl} \ , \eqno (85) $$ where $M_{Pl}$ is the
Planck mass, while;
$$ \Gamma = (1/4 \pi) \sqrt {45 / (\pi \varsigma )} \ , \eqno (86) $$
and $\varsigma$ is the effective number of different spin states of
relativistic particles. Using this and defining $\tau_Q = 1 / \dot
\epsilon $ at the instant when the critical temperature is reached we
get:
$$ \tau_Q = 2 \Gamma M_{Pl} / T_C^2 \ . \eqno (87) $$ We can also
estimate, by using a (weak coupling) approximate equality;
$$ T_C = \sqrt {\alpha (T=0)/\beta } = \sqrt {{\alpha '}/\beta} \ ,
\eqno (88)$$ so that:
$$ \tau_0 = 1/(\sqrt {\beta} T_C) \ . \eqno (89) $$ Consequently,
$$ \hat t = (\beta^{\frac 1 3 } T_C)^{-1} (2 \Gamma M_{Pl} /
T_C)^{\frac 1 3 } \ , \eqno (90) $$ and;
$$\epsilon (\hat t) = (T_C / (2 \Gamma M_{Pl}))^{\frac 2 3} / \beta
^{\frac 1 3} \ . \eqno (91) $$ By contrast, the relative temperature
corresponding to the Ginzburg condition, Eq. (45), is simply given by:
$$ \epsilon _G = 2 \beta \ . \eqno (92)$$ Here, we have again used the
approximate relation, Eq. (88), as well as Eq. (9) of Ref. 2.

Thus, it appears that the initial density of defects expected on the
basis of the freezeout scenario (which seems to have been borne out in
the superfluid helium Lancaster experiment$^{8,9}$) is quite different
from the one anticipated on the basis of the activation (Ginzburg)
version of the original Kibble mechanism$^2$. This difference will be
especially pronounced in the phase transitions occurring at lower
energies, where $T_C / M_{Pl} \ll 1$.  Examination of the cosmological
consequences of this result is beyond the scope of this paper.

\bigskip
\noindent
{\bf 6.\quad ACKNOWLEDGEMENTS}\\

Discussions with Andreas Albrecht, Robert Brandenberger, Ike Chuang,
Anne Davis, Alasdair Gill, Nigel Goldenfeld, Tom Kibble, Bernard Yurke, Alex
Vilenkin, Wojtek Zakrzewski and other participants of the
{\it Topological Defects Programme} are gratefully acknowledged.  I
have especially benefited from the interactions with Peter McClintock
and Richard Lee concerning the superfluid helium experiment they (and
their colleagues) have carried out.  Isaac Newton Institute provided a
partial support and a
stimulating environment for much of the research reported here.
Discussion of Section 4. of this paper has been influenced by the
collaboration with Pablo Laguna aimed at the numerical study of the
evolution of the order parameter, and carried out with the support of
NASA HPCC ``Grand Challenge'' initiative.

\bigskip

\vfill
\eject

\noindent
{\bf 7.\quad REFERENCES}
\begin{quote}
\item 1. Ya. B. Zeldovich, I. Yu. Kobzarev, and L. B. Okun, Cosmological
consequences of a spontaneous breakdown of discrete symmetry, {\it
Sov. Phys. JETP}, 40:1-5 (1975).

\item 2. T. W. B. Kibble, Topology of cosmic domains and strings,
{\it J. Phys. A: Math. Gen.} 9:1387-1398 (1976); also,
T. W. B. Kibble, in this volume.

\item 3. A. Vilenkin and E. P. S. Shellard, {\it Cosmic Strings and other
Topological Defects} (Cambridge Univ. Press, Cambridge, 1994).

\item 4. W. H. Zurek, p. 479 in {\it Proc. DPF Meeting of APS}, T. Goldman and
M. M. Nieto, eds., (World Scientific, Singapore, 1985); W. H. Zurek,
Experimental cosmology: Strings in superfluid helium, Los Alamos
preprint LAUR 84-3818 (1984).

\item 5. W. H. Zurek, Cosmological experiments in superfluid helium?
{\it Nature} 317: 505-507 (1985).

\item 6. I. Chuang, R. Durrer, N. Turok, and B. Yurke, Cosmology in the
laboratory: Defect dynamics in liquid crystals, {\it Science}
251:1336-1338 (1991).

\item 7. B. Yurke, Coarsening dynamics in liquid crystal systems,
this volume.

\item 8. P. C. Hendry, N. S. Lawson, R. A. M. Lee, P. V. E. McClintock,
and C. H. D. Williams, Generation of defects in superfluid He$^4$ as
an analogue of the formation of cosmic strings, {\it Nature}
368:315-317 (1994).

\item 9. P. C. Hendry, N. S. Lawson, R. A. M. Lee, P. V. E. McClintock,
and C. H. D. Williams, Cosmological experiments in He$^4$ -- problems
and prospects, this volume.

\item 10. M. J. Bowick, L. Chandar, E. A. Schiff, and A. M. Srivastava,
The cosmological Kibble mechanism in the laboratory: String formation
in liquid crystals, {\it Science} 263:943-945 (1994).

\item 11. D. R. Tilley and J. Tilley,
{\it Superfluidity and Superconductivity}, second edition (Hilger,
Boston, 1986).

\item 12. M. Tinkham, {\it Introduction to Superconductivity},
(McGraw-Hill, New York, 1975).

\item 13. W. H. Zurek, Cosmic strings in laboratory superfluids and
topological remnants of other phase transitions, {\it Acta Physica
Polonica} 24:1301-1311 (1993).

\item 14. E. A. Lynton, {\it Superconductivity}, third edition,
(Menuthen, London, 1969).

\item 15. J. Tate, B. Cabrera, and S. B. Felch, Novel noise thermometer
for measuring the local critical temperature of a superconducting
ring, in {\it LT-17}, U. Eckern, A. Schmid, W. Weber, and H. W\"uhl,
eds.  (Elsevier, Amsterdam, 1984).

\item 16. S. Rudaz and A. M. Srivastava, {\it Mod. Phys. Lett.} A8:1443-1445
(1993).

\item 17. J. M. Blatt, Persistent ring currents in an ideal Bose gas,
{\it Phys. Rev. Lett.} 7:82-83 (1961).

\item 18. J. Bardeen, Quantization of flux in superconducting cylinder,
{\it Phys. Rev. Lett.} 7:162-163 (1961).

\item 19. J. B. Keller and B. Zumino, Quantization of fluxoid in
superconductivity, {\it Phys. Rev. Lett.} 7:164-165 (1961).

\item 20. P. Laguna and W. H. Zurek, work in progress.
\end{quote}
\end{document}